\documentstyle[12pt,epsf]{article}    
\def\@biblabel#1{[#1]\hfill}

\begin{document}           

\baselineskip 14pt plus 2pt

\newcommand{\npi}{\hspace{-0.5cm}} 
\newcommand{\smq}{\mbox{$\simeq$}}
\newcommand{\prl}[1]{Phys. Rev. Lett. {\bf {#1}}}
\newcommand{\prb}[1]{Phys. Rev. {\bf B {#1}}}
\newcommand{\fg}[1]{Fig.~\ref{#1}}

\npi {\bf
 Multi-Exciton Spectroscopy of a Single Self Assembled Quantum Dot
}\\ \\
{\bf E. Dekel$^a$, D. Gershoni$^a$, E. Ehrenfreund$^a$, D. Spektor$^a$}
{\bf J.M. Garcia$^b$ and P.M. Petroff~$^b$ }\\
$^a$Physics Department and Solid State Institute, Technion-Israel Institute
of Technology, Haifa, Israel\\
$^b$Materials Department, University of California, Santa Barbara, CA 93106 \\
\\
\rightline{\today}

\npi{\bf Abstract }\\
We apply low temperature confocal optical microscopy
to spatially resolve, and spectroscopically study
a {\bf single} self assembled quantum dot. 
By comparing the emission spectra obtained at various excitation 
levels to a theoretical many body model, we show that:
Single exciton radiative recombination is very weak.
Sharp spectral lines are due to optical transitions between confined
multiexcitonic states among which excitons thermalize within their
lifetime. Once these few states are fully occupied, broad bands
appear due to transitions between states which contain continuum electrons.

\newpage

The study of electronic processes in semiconductor heterostructures 
of reduced dimensionality has been a subject of recent extensive research 
efforts. Of particular importance are the efforts to fabricate and 
study semiconductor quantum dots (QDs) of nanometer size, 
in which the charge carriers are confined in all directions to 
characteristic lengths which are smaller than their De-Broglie wavelengths
[1--16]. 
These efforts are motivated by both the QDs potential device applications,
as well as their being an excellent stage for experimental studies of basic
quantum mechanical principles. 
One very promising system of such QDs is called
self assembled QDs (SAQDs). In fabricating SAQDs, minimization of the 
lattice mismatch strain between different epitaxially grown semiconductor 
layers occurs via the formation of small islands connected by a very 
thin wetting layer. By capping these self assembled islands 
with an epitaxial layer of wider bandgap material, with similar
lattice constant to that of the substrate, high quality QDs are 
produced \cite{marzin,drexler}. This natural way of producing large 
ensembles of QDs has motivated a vast number of studies of their structural,
electronic and optical properties [3-5,10-16].
The size distribution of these SAQDs (typically about 10\%), and the resultant
inhomogeneous broadening of the SAQDs characteristic features, has 
so far limited the ability to clearly understand and unambiguously interpret
the experimental results.
In this Letter, we overcome this obstacle by spectroscopically
studying multi-excitonic optical transitions in a {\bf single}  SAQD.
We show here, indeed, that multiple sharp spectral lines, as well as broad
spectral features, which previously were interpreted as an optical
signature for emission from an ensemble of dots \cite{marzin,drexler}, 
are actually due to optical transitions between multi carrier states 
within a single dot, under various excitation levels.

The  SAQD sample studied here was fabricated by 
deposition of a coherently strained epitaxial layer of $InAs$ on an $AlGaAs$ 
layer deposited on $GaAs$ substrate. 
The layer sequence, compositions and widths are given in the left inset to
\fg{scan}.
During the growth of the strained layer, the sample was not rotated, 
thus a gradient in the QDs density was formed across its 
surface \cite{fafard}. In particular, low density areas,
in which the average distance between neighboring QDs is
larger than our spatial resolution could easily be found on the sample surface.
The right side inset to \fg{scan} displays the far-field ensemble
photoluminescence (PL) spectrum of such an area of the SAQD sample.

We use a $\times 100$ in-situ microscope objective for diffraction limited
low temperature confocal optical spectroscopical studies of
the single SAQDs. Our system provides spatial resolution of 
\smq0.5 $\mu m$, both in the excitation and the detection channels \cite{erez}.

The dots position and characteristic emission wavelength are found by 
taking PL spectra during a line scan over the SAQD sample surface.
A typical scan is displayed in \fg{scan}, where the PL intensity as 
a function of photon energy and objective position is given by the 
gray color scale. 
During this scan, the PL was excited with a 730 nm, $7 \mu W$ cw light
from a Ti:S laser.  In each 0.1 $\mu m$ long step of the objective the
PL spectrum was measured by exposing a cooled CCD camera for 50 sec. 
Three emission lines from three different spatial 
positions along the scanned line are evident in \fg{scan}. 
These lines are due to carriers' recombination within 
single SAQDs, as indicated by their spatial and spectral widths, 
which are both resolution limited \cite{gammon,hess}.
In \fg{pl} we present PL spectra from a single SAQD found by such a scan
for various excitation powers. 
An overall perception of the important spectral features 
is obtained from the 100 $\mu$W spectrum which is composed of two groups of 
emission lines 
located near 1.325 and 1.375 eV, respectively. The groups are nearly 
symmetrically positioned  around a weak emission line at 1.355 eV. 
Each group is composed of several sharp and well resolved lines, 
two of which, roughly 7 meV apart, are particularly strong. 
We marked the strongest lines in each group and the center line by numbers 
in increasing order of their spectral position.
At excitation power of 1$\mu$W, only lines 1 and 3 are observed, at intensity
of less than one count per second, in agreement with the estimated exciton 
lifetime \cite{raymond} and our system's collection efficiency.
The emission of line 1 increases roughly as the square root of the
excitation power up to about 100 $\mu W $, where it reaches saturation.
Line 3, on the other hand, reaches saturation already at excitation power
of 7 $\mu$W. 
At 20 $\mu W$ the emission spectrum is already composed of the five 
main spectral lines, 
and at 100 $\mu W$, all main lines are saturated. 
At saturation, the intensity of line 2 is about half that of lines
1, 4 and 5, whose intensity is \smq10 times larger than that of line 3.
Above 100 $\mu$W, several sharp lines appear below each group of lines. 
At yet higher excitation power, these lines form two broad 
spectral bands (C1 and C2, Fig. 2) which dominate the PL spectrum. 

For the analysis of our observations we use a model
parallelepipedal box with infinite potential barriers and rectangular 
base whose dimensions are much larger than its height.
We fitted the dot base dimensions to obtain the observed level
separation of \smq50 meV and adjusted its height to obtain the Coulomb 
splitting of 7 meV (see below). 
Using typical InAs effective masses \cite{adachi} for electrons (0.023 $m_0$)
and for heavy holes (0.6 $m_0$, where $m_0$ is the electron rest mass)
and dielectric constant \cite{adachi} ($\varepsilon=15$) 
we find that base dimensions 
of 30.2$\times$31.2 nm$^2$ and height of 5 nm best fitted the 
observed data. 
Note that the box base is not square, thus there is no geometrical 
degeneracy in agreement with recent calculations\cite{zunger2}. 
Though our model does not describe the exact SAQD potential structure
in geometrical shape \cite{drexler}, strain and  
piezoelectric fields \cite{pryor,zunger2}, it is reassuring that the fitted
dimensions are similar to those typically reported for this SAQD system \cite{drexler}. 
We show below that in spite of its simplicity the 
model explains our data quite well. This means that the knowledge of the
single particle level separation and the strength of the Coulomb matrix 
elements between these levels are nearly enough to describe the optical
properties of a fully quantized system with a few carriers in it. 
The details of its confining potentials are only second in importance. 

We consider the first
four, doubly degenerate electron and hole levels in our model dot:
(111), (121), (211) and (221) at this energy order, where the numbers in
parentheses are the quantum numbers associated with the confinement along
the cartesian axes x, y and z (the growth direction), respectively. 
The wavefunctions of electrons and holes in these states 
are analytically expressed. 
This limited number of states is adequate to explain the low level 
excitation PL spectra. Higher excitation levels, which give rise 
to carriers in the continuum above the dot barriers' potential, 
are dealt here more qualitatively. 
The Coulomb interaction between carriers in our dot is now 
considered using the following many body Hamiltonian:
$H=H_{free}+H_{Coul}~,$
where $H_{free}$ describes non interacting electrons (e) and holes (h)
in their respective bands, and $H_{Coul}$ describes the e-e, h-h and e-h
Coulomb direct and exchange interactions \cite{bd}.
Since electron and hole wavefunctions are identical in our model,
the e-h exchange interaction term vanishes.
We find the solution for the multiexciton energy 
levels and wavefunctions, by exact diagonalization of the
the many-body Hamiltonian for these first four single carrier levels. 
We consider here {\bf all} neutral
multi-excitonic states up to exciton population number of eight,
at which all the levels considered are fully occupied.
Optical transitions between the different excitonic population levels, 
in which a single e-h pair is annihilated, are then
calculated using the dipole approximation \cite{bd}.

In \fg{levels}, horizontal solid lines represent some of the calculated 
energy levels. The single carrier states (for which the Coulomb 
interaction is ignored) are displayed on the left side 
of the figure. Single electron (hole) energy levels are represented by
superscripts above the letters e(h) and the occupation number of these levels
are given by numbers in front of the letter. The state degeneracy (due to
the carriers' spin) is given by the number in parenthesis.
The multi-excitonic states, including now the Coulomb interaction between
the carriers, are displayed on the right side of the figure.
Here, the degeneracy is partially removed and we describe the states by
their total angular momentum (S), its projection on the dots growth
direction ($M_S$), and the total electronic ($S_e$) and hole
($S_h$) spins.
With one exception, indicated by  dashed horizontal lines, 
we display only those states which evolve from single carrier levels of
identical electron and hole quantum and occupation numbers.
This is because the lowest energy state in each excitonic occupation level
is always of this type, and thus, optical transitions between these levels 
dominate the PL spectrum.
These single carrier states are either 1, 4 or 16 times 
degenerate. The Coulomb interaction lifts only the degeneracy of 
the 16 multiplets, which occur whenever two half filled levels of 
electron and hole participate in the multiexcitonic state. They are split 
into 3 levels according to their $S_e$ and $S_h$ quantum numbers,
exactly like a bulk biexciton \cite{forney}.
The lowest of these levels is 9 fold degenerate containing 
a quintuplet, a triplet and a singlet of 
$S_e =1$ and $S_h = 1$ which add to $S$= 2,1 and 0, respectively.
The mid energy level is 6 fold degenerate 
containing two triplets of $S_e=1(0) $ and $S_h=0(1) $ which add to 
$S=1(1)$. The highest of these levels contains only a singlet 
with $S_e=0$, $S_h =0$ and $S=0$. 
For calculating the PL spectrum 
the distribution of excitons among their multi excitonic states should be known.
The relatively small number of observed PL lines lead us to safely conclude
that only the lowest multi excitonic energy state of each
exciton occupation number has a significant steady state population. 
This means that 
there is no phonon bottle-neck \cite{bottleneck} for exciton thermalization
and they reach thermal distribution faster than the radiative 
recombination occurs.

The optically allowed transitions between the lowest multi excitonic 
state of each exciton occupation level and the multi excitonic states of 
one less exciton occupation are represented in \fg{levels} by vertical arrows. 
Annihilations of $e^1h^1$ ($e^2h^2$) are represented by 
solid (empty) arrows, and annihilation of $e^2h^1$ 
is represented by a dash arrow. We note that optical transitions 
conserve S and  $M_S$ \cite{bd}.
Transitions of the same energy between different exciton occupation levels
are drawn along the same vertical thin dashed line. These vertical lines are 
numbered in increasing order of their energies, very much like
the experimentally measured PL lines as shown in \fg{f4} below.
In \fg{f4} we display the calculated dot emission spectra, in units of the
squared dipole moment between the $e^1$ and $h^1$ states, for a few 
exciton occupation numbers. In the calculations, we average over degenerate
initial states and sum over final states. For comparison, measured spectra
for various excitation powers are shown by dots.

The following conclusions are drawn by comparing the
experimental data to the model calculations.
The characteristic appearance of the PL lines in pairs,
is clearly explained in terms of the Coulomb splitting of biexciton
levels (the nine and six degenerate levels).
The multi line PL spectrum which we measure even at our lowest excitation
power, indicates that the average exciton occupation number is already
larger than two. This agrees very well with a conservative estimate
of the occupation number, based on the measured laser power, and estimations
of the absorption and exciton lifetime \cite{raymond}
(Diffusion of carriers from the AlGaAs layers into the SAQD, can be safely 
neglected due to the sample structure). 
Thus, surprisingly, single lines due to the recombination
of single exciton and single biexciton are not observed at all at 
low excitation spectra. Line 2, which corresponds in energy to the 
radiative recombination of a single
exciton, appears only at much higher excitation power (20 $\mu W $). 
This can be due to  e-h exchange,
which splits the four fold degeneracy between
the triplet and singlet states of the single exciton level.
Since, as has been observed in II-VI
nanocrystallites \cite{bawendi}, the lower energy triplet state
is optically forbidden, the single exciton annihilation line 
can only  be observed when at least three excitons occupy the dot.
Alternatively, it can be due to a reduced electron-hole 
overlap integral which weakens this transition.~\cite{zunger2}

Assuming that recombination is possible only via radiative 
channels we calculated the lifetime of each multi excitonic occupation level.
Using these lifetimes in turn, a correlation between the excitation power
and the number of excitons within the dots can be drawn.
We find a relative good agreement between the excitation power
dependence of the PL spectra and the calculated spectra based on that 
correlation. 
The appearance of sharp lines at the lower energy side of
each of the two spectral line groups, and their evolution with increasing
excitation power into the broad bands C1 and C2, are also explained
by our model (see the 500 $\mu W$ PL spectrum in \fg{f4}).
We note in \fg{levels} that the higher the excitation is,
the higher are the populated single electron and hole levels
which participate in the multi excitonic ground states.
Optical transitions from these states to the high energy states of
one less exciton level by the annihilation of $e^1h^1$ and $e^2h^2$ are now 
possible, leading to a characteristic decrease in the energy of these
transitions. The magnitude of this shift is
given by the Coulomb interaction term between the high energy level and
the lower one. This term is almost constant as long as confined 
single carrier states are concerned. At high enough excitation power, 
when continuum electron levels are populated, 
these shifts are becoming smaller and thus, broad spectral bands at low 
energy are formed. Since the density of continuum levels is large,
these bands are not saturated, and they eventually dominate the
PL spectra.
We do not observe higher energy PL lines due to the annihilation of higher 
energy e-h pairs, such as $e^3h^3$ and $e^4h^4$
probably since, as can be seen in \fg{pl}, they are masked by the large PL
emission from the GaAs substrate.

In summary, we resolved the emission from a single self assembled quantum dot
and successfully explained its power dependent PL spectra using multi carrier
Hamiltonian.


\newpage

\npi {\bf Figure Captions}

\newcounter{fg}
\refstepcounter{fg}
\label{scan}
\refstepcounter{fg}
\label{pl}
\refstepcounter{fg}
\label{levels}
\refstepcounter{fg}
\label{f4}
     
\begin{list}
{\fg{\arabic}}{\usecounter{fg}}

\item[\fg{scan}]
PL intensity as a function of photon energy and position
along a line across the SAQD sample surface.
The intensity is given by color scale bar to the right. 
Left inset: schematic description of the sample.
Right inset: far-field PL spectrum of the sample.
\item[\fg{pl}]
PL spectra from a single SAQD for various excitation power levels.
\item[\fg{levels}] 
Energy level diagram of the SAQD multiexcitonic states, contributing the most 
to optical emission. 
Shown are the multiexcitonic levels calculated by excluding (left) 
and including (right) the Coulomb interaction between carriers.
The vertical arrows indicate optical transitions due 
to one exciton annihilation, (see text).
\item[\fg{f4}]
Calculated PL spectra for various exciton population numbers.
For comparison, the experimental measurements are also displayed.
\end{list}

\epsfxsize=13.5cm
\epsffile{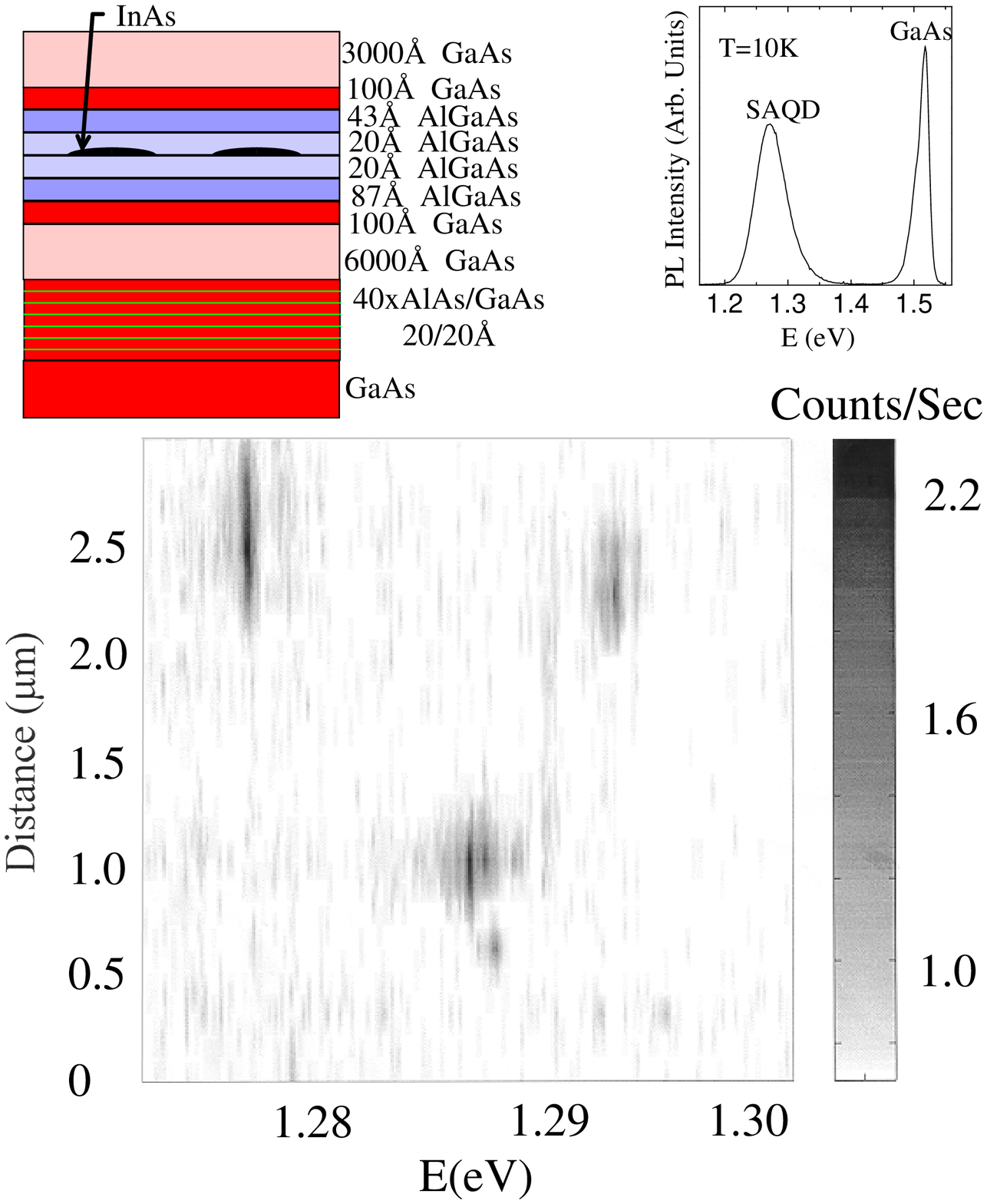}
\vspace {0.5cm}

Fig.1

\epsfxsize=13.5cm
\epsffile{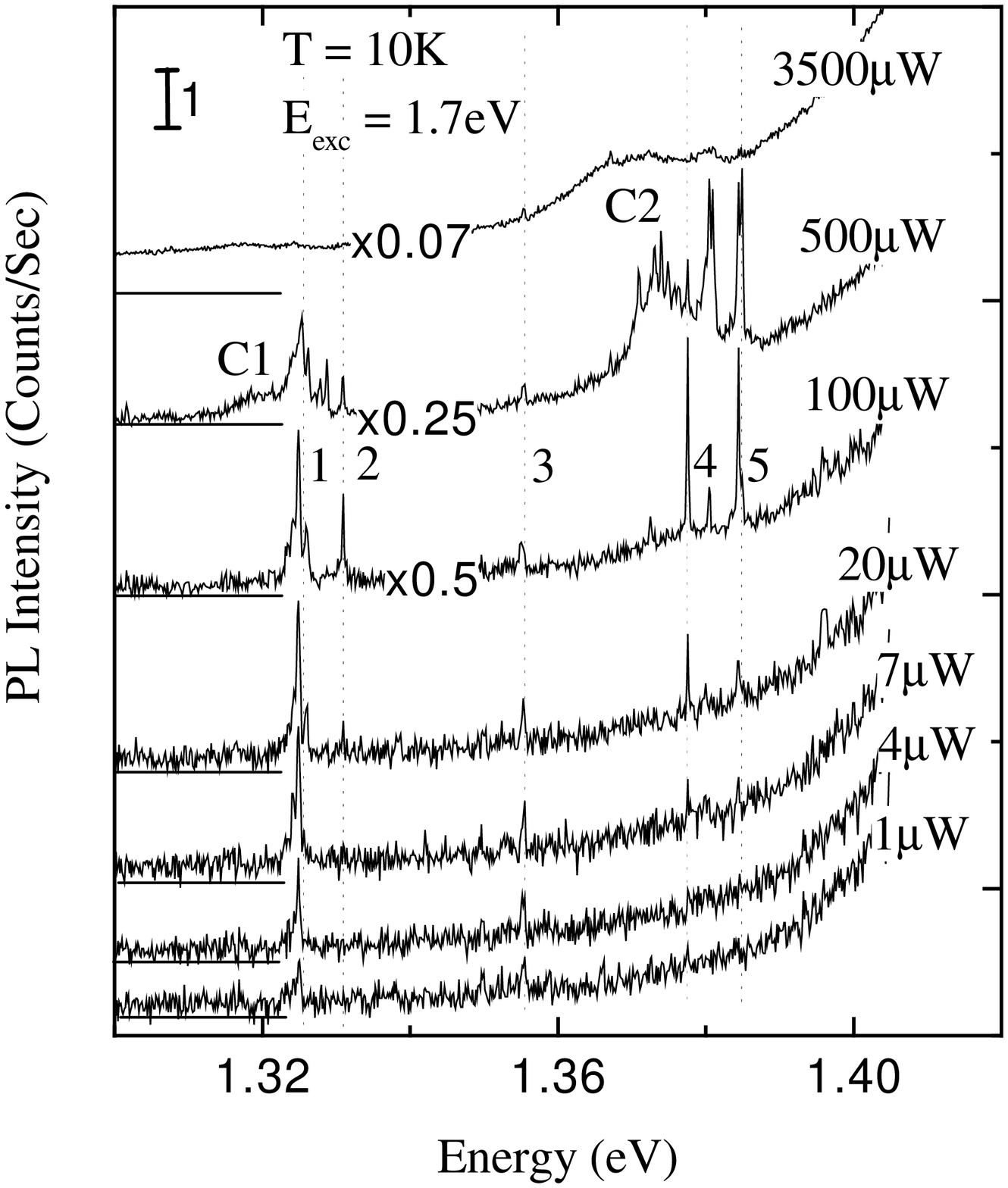}
\vspace {0.5cm}

Fig.2

\epsfxsize=13.5cm
\epsffile{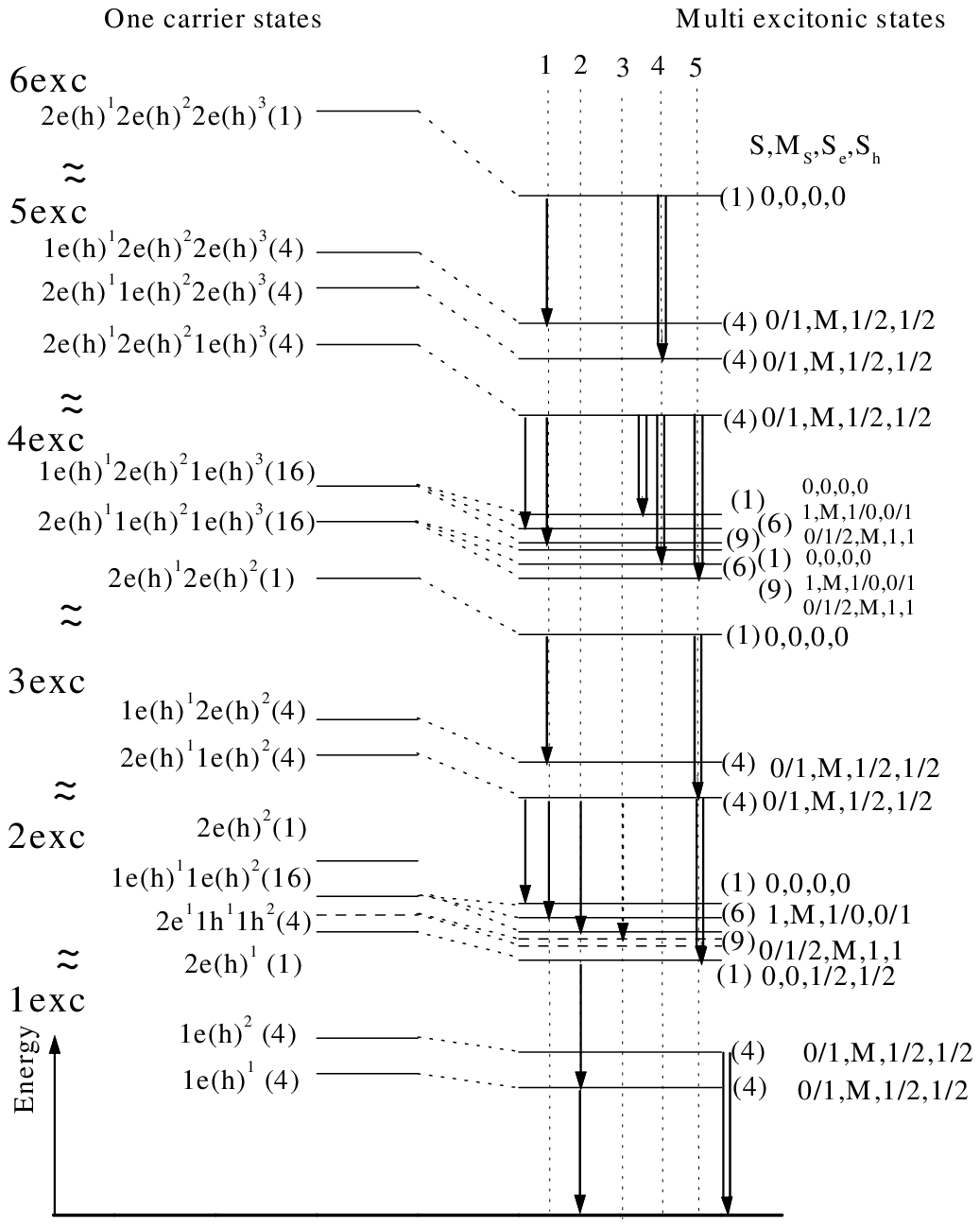}
\vspace {0.5cm}

Fig.3

\epsfxsize=12.5cm
\epsffile{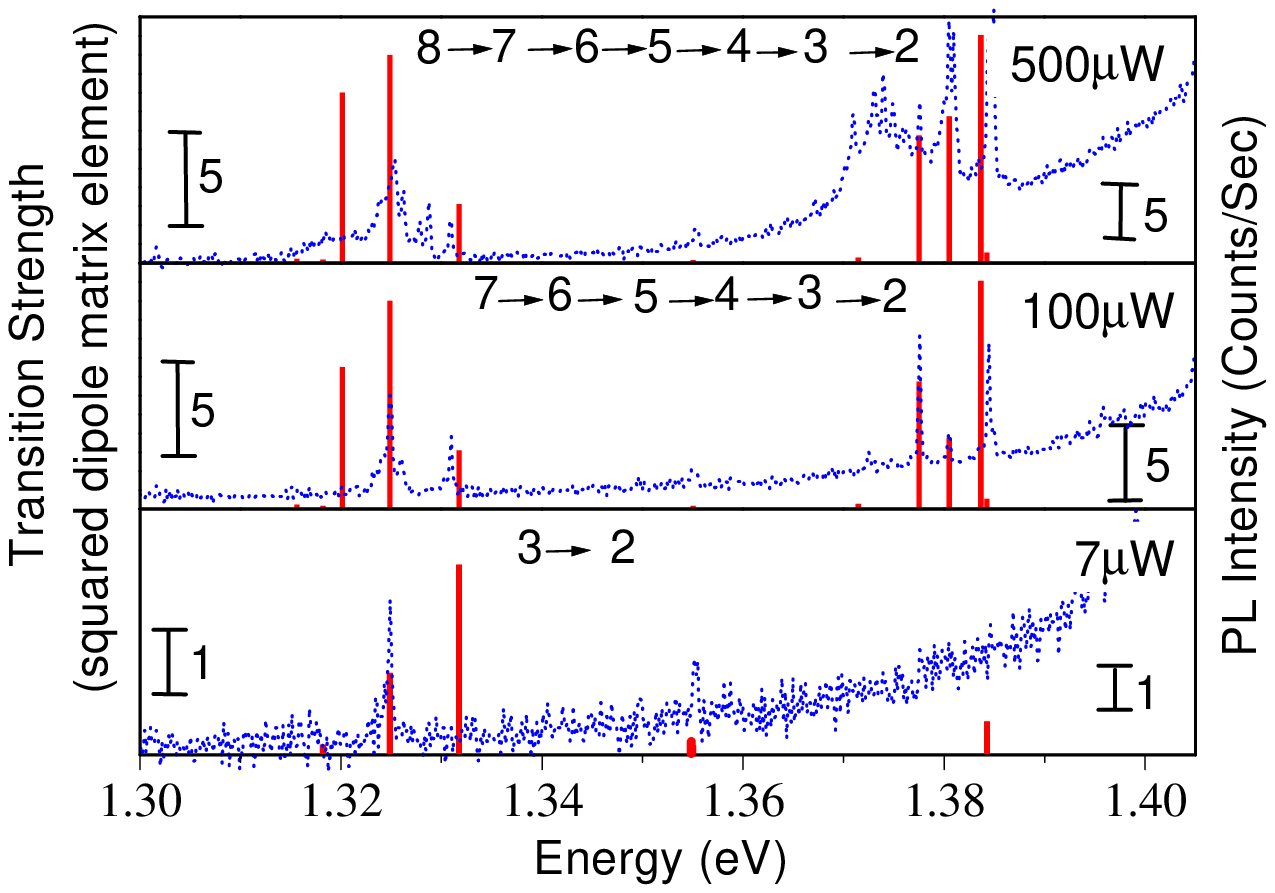}
\vspace {0.5cm}

Fig.4

\end{document}